\documentclass[prd,twocolumn,floatfix,preprintnumbers]{revtex4}

\usepackage{graphicx}
\usepackage{graphicx,epsfig}
\usepackage{bm}
\usepackage{latexsym,amssymb,amsmath,float}

\newcommand {\ga} {\ {\raise-.5ex\hbox{$\buildrel>\over\sim$}}\ }
\newcommand {\la} {\ {\raise-.5ex\hbox{$\buildrel<\over\sim$}}\ }

\def\be{\begin{equation}}
\def\ee{\end{equation}}
\def\ba{\begin{eqnarray}}
\def\ea{\end{eqnarray}}

\renewcommand{\(}{\left(}
\renewcommand{\)}{\right)}
\renewcommand{\[}{\left[}
\renewcommand{\]}{\right]}

\begin{document}

\title{The Coincidence Problem and the Swampland Conjectures in the
Ijjas-Steinhardt Cyclic Model of the Universe}
\author{Robert J. Scherrer}
\affiliation{Department of Physics and Astronomy, Vanderbilt University,
Nashville, TN  ~~37235}

\begin{abstract}
In the Ijjas-Steinhardt cyclic model, the universe
passes through phases dominated by radiation, matter,
and a dark energy scalar field, with the value of the scale
factor increasing with each cycle.  Since each cycle terminates
in a finite time, it is straightforward to calculate the fraction of time that
the universe
spends in a state for which the matter and dark energy
densities have comparable magnitudes; when this fraction is large,
it can be taken as a solution of the coincidence problem.
This solution of the coincidence problem requires a relatively short lifetime for each cycle, but
unlike in the case of phantom models, there is no fixed upper bound
on this lifetime.  However, scalar field models
satisfying the Swampland conjectures yield sufficiently
short lifetimes to provide a satisfactory resolution of the coincidence problem.
\end{abstract}

\maketitle

Cosmological data \cite{union08,hicken,Amanullah,Union2,Hinshaw,Ade,Betoule}
indicate that roughly
70\% of the energy density in the
universe is in the form of a negative-pressure component,
called dark energy, with approximately 30\% in the form of nonrelativistic matter (including both baryons
and dark matter).
The dark energy component is often parametrized by its equation of state parameter, $w$,
taken to be the ratio of the dark energy pressure to its density:
\be
\label{w}
w=p_{\rm DE}/\rho_{\rm DE},
\ee
where $w=-1$ corresponds to a cosmological constant.
If $w$ is roughly constant, the density of the dark energy, $\rho_{DE}$, scales as
\be
\label{rhode}
\rho_{DE}=\rho_{DE0} (a/a_0)^{-3(1+w)},
\ee
where $a$ is the scale factor, and $\rho_{DE0}$ and $a_0$ are the dark energy
density and scale factor, respectively, at the present.
(We will use zero subscripts throughout to refer to present-day values). 
Current observations constrain $w$ to be relatively close to $-1$.

The matter density, in contrast, scales as
\be
\label{rhom}
\rho_M = \rho_{M0} (a/a_0)^{-3}.
\ee
This leads to a problem:  while the matter and dark energy densities
today are within nearly a factor of two of each other, at early times, $\rho_M \gg \rho_{DE}$,
and in the far future we expect $\rho_{DE} \gg \rho_M$.  It would appear, then, that we live in a 
very special time:  this is the well-known coincidence problem.

Many solutions have been proposed for the coincidence problem, but we will concentrate here on
one particular approach: models in which the universe
can be shown to spend a significant fraction of its lifetime in a ``coincidental"
state.  In Ref. \cite{Scherrer}
it was suggested that the coincidence
problem could be resolved in the context of phantom dark energy models.
In such models, $w < -1$, and the universe terminates in a singularity at a finite time \cite{Caldwell2,Caldwell3},
so that the fraction of time for which the dark energy and matter densites are relatively close
can be a significant fraction of the universe's (finite) lifetime.

This result
was extended to
phantom models with a time-varying equation of state in Ref. \cite{Kujat} and
to scalar field models with a linear potential in Ref. \cite{Avelino}.
In the latter models, the dark energy density can evolve toward negative values, causing the universe to cease
expanding and recollapse into a final singularity \cite{Felder,Kallosh}, so the universe
has a finite lifetime and can spend a significant fraction of that lifetime in a state
with roughly equal abundances of matter and dark energy.  Both the linear models  and
the phantom dark energy model were further examined as solutions to the coincidence problem 
in Ref. \cite{BA}.

Applications of this approach to cyclic phantom models were explored by Chang and Scherrer
\cite{CS}, who examined the particular model of
Ilie et al. \cite{Ilie}. (See also similar models in Refs. \cite{Crem,Xiong}
and the much earlier model of Ref. \cite{BD}, a more
conventional oscillating model with positive curvature and a cosmological constant.  While Ref. \cite{BD}
does not attempt to solve the coincidence problem, it does make an argument
for the ``largeness" of the cosmological constant relative to the radiation/matter components).
In cyclic phantom models,
the universe goes through repeated
cycles of matter/radiation domination followed by a phantom dark energy phase.
Within
each cycle, there is a significant period in which the dark energy
and matter densities are comparable.  Since these cycles repeat endlessly,
it is not surprising that we find ourselves in an epoch in which
the dark energy and matter densities are of the same order of magnitude.

Here we propose a similar argument in the context of the cylic model recently put forward
by Ijjas and Steinhardt \cite{IS}.  In the Ijjas-Steinhardt (IS) model, the universe
undergoes alternating periods of expansion and much shorter periods of contraction.  While the overall scale factor
increases with each cycle, the relative values of $a$ within a single cycle remain unchanged
from one cycle to the next, so the universe appears to repeat the same expansion behavior
within each cycle.  This model has several interesting features.  It resolves
many of the same problems as inflation, such as the monopole, flatness, and horizon problems,
and the evolution of the universe remains purely classical at all times, without an initial
or final singularity.

Conceptually, the discussion of the coincidence problem in the IS model most closely
resembles that of Ref. \cite{CS} for cyclic
phantom models.  However, the IS model differs significantly from these models
in lacking a final singular state.  The dark energy dominated state in the IS model
can be arbitrarily long, making the fraction of time spent in a state with roughly
equal matter and dark energy densities very small.  We will show that this problem can be remedied in the context
of the Swampland conjectures, which force a relatively early termination of the expanding
phase of the universe relative to the present day.

In the IS model, the universe
contains the standard components of radiation, nonrelativistic matter (baryons and dark matter)
and dark energy, which is assumed to
be in the form of a scalar field (quintessence).
The dark energy drives the accelerated expansion of the universe at late
times, but the universe eventually undergoes a transition to an
ekpyrotic contracting phase, driven by the same scalar field.  This contracting phase
lasts a very brief time compared to the expanding phase.  At the conclusion of
the contracting phase, the epkyrotic field is converted into matter and radiation, and
another cycle begins.

Because the evolution of the universe is cyclic, with a finite lifetime for each cycle,
it is possible to calculate the fraction of time that the universe spends in a coincidental
state, defined to be a state for which the ratio of the density of dark energy to the density
of matter lies within some fixed range close to 1.  Note that the periodicity of
the universe is crucial in this argument; a universe undergoing a single expansion to
a final de Sitter phase spends an infinite time with $\rho_{DE} \gg \rho_{M}$, and
the fraction of its lifetime spent in a coincidental state is effectively
zero (to the extent that this fraction is defined at all).  While not strictly required by the IS model, we will take the final
expanding state to be close to de Sitter, with $w_{DE} \approx -1$, in agreement
with current observational data.

Let $\rho_{DE}$ be the dark energy density, and $\rho_{M}$ be the nonrelativistic
matter density, and define the coincidence ratio $r$ as in Ref. \cite{Scherrer}:
\begin{equation}
\label{rdef}
r \equiv \frac{\rho_{DE}}{\rho_{M}} = \frac{\rho_{DE0}}{\rho_{M0}}\(\frac{a}{a_0}\)^3.
\end{equation}
We will then define a coincidental state to be one for which $r$ lies
sufficiently close to one, where the definition of ``sufficiently close"
is, of course, somewhat arbitrary.

We assume a flat Friedman-Robertson-Walker model
and work in reduced Planck units ($\hbar = c = 8\pi G = 1$),
so that the
evolution of the scale factor is given by
\be
\label{Fried}
\(\frac{\dot {a}}{a}\)^{2}=\frac{\rho}{3}.
\ee
The era of radiation domination constitutes a minuscule fraction
of the lifetime of each cycle, so we can neglect it and consider
only the time during which matter and dark energy are dominant, with
the density of the latter assumed to be nearly constant.
Then Eq. (\ref{Fried}) becomes
\be
\(\frac{\dot {a}}{a}\)^{2}=\frac{1}{3}\[\rho_{M0}\(\frac{a}{a_0}\)^{-3}+\rho_{DE0}\].
\ee
In terms of the coincidence ratio $r$ defined in Eq. (\ref{rdef}), we have
\be
\(\frac{1}{3} \frac{\dot r}{r}\)^2 = \frac{1}{3}\rho_{DE0}\(\frac{1}{r} + 1 \)
\ee
This equation can be integrated exactly to give
\begin{equation}
\sinh^{-1}\sqrt{r} = \sqrt{(3/4)\rho_{DE0}} ~t.
\end{equation}

Now we can calculate the fraction of each cycle that the universe
spends in a coincidental state, with $r$ near 1.  As already noted,
we neglect the radiation-dominated portion of each cycle.  We will
also neglect the time spent in the contracting phase, as this is
assumed to be much smaller than the time over which the universe
is expanding \cite{IS}.  Since each cycle in the IS model is identical to
all of the others, modulo an overall expansion factor, we can simply
calculate the coincidence fraction for the current cycle.  Following
Ref. \cite{Scherrer}, we will
define a coincidental state to correspond to
\be
r_1 < r < r_2.
\ee
Then the fraction of time $f$ that the universe spends
in this coincidental state is
\be
\label{fdef}
f = \frac{\sinh^{-1}\sqrt{r_2} - \sinh^{-1}\sqrt{r_1}}{\sqrt{(3/4)
\rho_{DE0}}~t_{cyc}},
\ee
where $t_{cyc}$ is the lifetime of the current cycle, i.e., the time at which
expansion ceases and contraction commences.  Since the cycles
repeat indefinitely, $f$ is also the fraction of the entire lifetime of the universe
that is spent in a coincidental state.  We can rewrite Eq. (\ref{fdef}) in terms of the 
present day ratio of dark energy density to dark matter density, $r_0$, and
the present age
of the universe, $t_0$:
\be
\label{ffinal}
f = \frac{\sinh^{-1}\sqrt{r_2} - \sinh^{-1}\sqrt{r_1}}{\sinh^{-1}\sqrt{r_0}}
\(\frac{t_0}{t_{cyc}}\).
\ee
The choices for $r_1$ and $r_2$ are somewhat arbitrary.  Here we will define the state
of the universe to be ``coincidental" if the matter and dark energy densities are within
an order of magnitude of each other, i.e., $r_1 = 1/10$ and $r_2 = 10$.  Then
taking $r_0 \approx 7/3$, we obtain
\be
\label{f}
f = 1.3\(\frac{t_0}{t_{cyc}}\).
\ee
Note that the derivation of Eqs. (\ref{fdef})-(\ref{f}) assumes that
$t_{cyc} > t_2$, where $t_2$ is the value of $t$ when $r = r_2$.
For our choice of $r_2$, this corresponds to the limit
$t_{cyc}/t_0 > 1.9$, so $f < 0.7$.  In what follows, this limit will always be satisfied.

As we would expect, $f$ varies inversely with $t_{cyc}$.  As an illustrative
example, Ijjas and Steinhardt discuss the case where $t_{cyc} = 10t_0$, which
gives $f = 0.13$, so the universe spends 13\% of its time
in a state for which the matter and dark energy densities are within an order of
magnitude of each other.  For this particular case, $f$ is not so large that
the argument for this solution of the coincidence problem is compelling, but neither
is $f$ so absurdly small that the argument fails completely.

One might hope to invert this argument to make a Bayesian estimate of the likelihood
of a particular value of $t_{cyc}$ based on the fact that we do happen to observe
$\rho_{M} \sim \rho_{DE}$; such an argument would tend to favor relatively smaller
values of $t_{cyc}/t_0$.  (For an example of such arguments, see Ref. \cite{BA}).
The problem arises because the IS model does not include a prescription for
$t_{cyc}$, so $t_{cyc}$ could, in principle, be arbitrarily large.  It then becomes
impossible to define a reasonable prior on the distribution of $t_{cyc}$.

As an alternative, we can place an upper bound on $t_{cyc}$ by requiring the IS
model to satisfy the Swampland conjectures.  These conjectures
arise in the context of attempts to derive a quantum theory of gravity within string theory.
A variety of general constraints on such models have
been proposed (see, e.g., Refs. \cite{Arkani,OV} for some of the earliest work in this area).
Here we consider specifically the constraints examined in
Refs. \cite{Brandenberger,limit1,Akrami,limit2,Garg}.  Because these references
discuss the motivations for these conjectures in detail, here
we will simply summarize them.

Consider a scalar field $\phi$, with potential $V(\phi)$, that serves as the dark energy.
The Swampland conjectures we consider here provide an upper bound on the total distance $\Delta \phi$ over which $\phi$
can evolve, and a lower bound on the logarithmic derivative of $V$ with respect to $\phi$.
Specifically, they argue for the existence of constants $d$ and $c$, both of order unity, such that,
in reduced Planck units,

\noindent{Conjecture 1:}
\be
\label{swamp1}
\Delta \phi < d \sim \mathcal{O}(1),
\ee

\noindent{Conjecture 2:}
\be
\label{swamp2}
\lambda \equiv |V^\prime|/V > c \sim \mathcal{O}(1),
\ee
\noindent where $V^\prime$ is the derivative of $V$ with respect to $\phi$.

Conjecture 1 is longstanding \cite{OV} and has a great deal
of theoretical support (although see Ref. \cite{Scalisi} for mechanisms allowing
$d$ to be somewhat larger than $\sim \mathcal{O}(1)$).
Conjecture 2 is
more recent \cite{Obied}.  Conjecture 2 is inconsistent with standard $\Lambda$CDM and is in tension even with quintessence
models.  The problem arises because observations favor $w$ near $-1$ at moderate redshifts, but $w$ near $-1$ generally translates
into values of $\lambda$ less than 1.  Agrawal et al. \cite{limit1} show that for any quintessence model consistent
with observations, $\lambda$ can be at most $0.6$, implying that $c \la 0.6$.  Akrami et al. \cite{Akrami}
and Raveri, Hu, and Sethi \cite{limit2} find similar
limits on this parameter: $c < 0.54$ and $c < 0.51$, respectively, at the 95\% confidence level.
As the value of $c$ in the second Swampland conjecture is not specified exactly, it is fair to say that
quintessence models
are in moderate tension with this conjecture but not absolutely inconsistent at this point.

Now consider the implications of these Swampland conjectures for the IS model.
If we express the IS model in terms of ``thawing" quintessence models \cite{CL}, in which the scalar field
is initially at rest in a potential $V(\phi)$ with $w$ near $-1$ and slowly rolls down the potential, then the onset
of rolling is determined by the value of $\lambda$:  larger $\lambda$ corresponds to an earlier transition to rolling
behavior.  This is easy to see if we examine the quintessence equation of motion:
\be
\label{qevol}
\ddot{\phi}+ 3H\dot{\phi} + V^\prime =0,
\ee
where the dot denotes time derivative, and $H \equiv \dot a/a$.  For quintessence evolution (in which both the scalar field and matter densities
enter into the value of $H$), all three terms in Eq. (\ref{qevol}) are of the same order of magnitude (unlike the case
of slow-roll evolution in inflation).  Then at early times, the distance travelled by the field $\phi$ is
\begin{equation}
\Delta \phi \sim \ddot{\phi} t^2 \sim V^\prime t^2.
\end{equation}
Then using the fact that $t^2 \sim 1/(\rho_M + \rho_\phi)$, and the quintessence energy density is initially $\rho_\phi = V$,
we have
\be
\label{deltaphi}
\Delta \phi \sim \lambda \Omega_\phi,
\ee
where $\Omega_\phi$ is the fraction of the total density in the form of quintessence.  (Ref. \cite{limit1}
provides a more exact derivation of $\Delta \phi$ for the case of an exponential potential.)  Eq.
(\ref{deltaphi})
illustrates the constraints that the Swampland conjectures place on the IS model.
Conjecture 2 places a lower bound on $\lambda$ which, from Eq. (\ref{deltaphi}), gives a lower bound
on $\Delta \phi$.  Then
the de Sitter expansion phase
cannot continue indefinitely because the shape of the potential must change before $\Delta \phi$ violates Conjecture 1.

To say anything more quantitative requires a particular model for the quintessence field driving
the expansion phase of the IS model.  As a specific example, we will examine the quintessence
model with an exponential
potential:
\be
V(\phi) = V_0 \exp(-\lambda \phi),
\ee
with $\phi = \dot \phi =0$ at $t=0$.  (Note that the value of $V_0$ is arbitrary,
as it can be absorbed into a translation of $\phi$).  In this model, $\phi$ begins frozen at $\phi = 0$, but eventually
rolls down the potential, so that $w$ increases from $-1$ as the field thaws.
Quintessence with an exponential potential has been widely studied \cite{Wetterich,RatraPeebles,
FerreiraJoyce1998,CLW,Liddle}, but it is particularly important
with regard to the Swampland conjectures, because, as noted in Ref. \cite{limit1}, it allows for the largest
value of $V^\prime/V$ for thawing quintessence consistent with current observations.

In order to serve
as a quintessence field in the IS model, $V(\phi)$ must change sign at some value of $\phi$.
Hence, we will assume that $V(\phi)$ evolves sharply away from an exponential potential and
takes on a negative value
after $\phi$ has evolved over a distance $\Delta \phi$.  At this point, the universe begins
its contracting phase.

We have numerically integrated Eq. (\ref{qevol}) for a range of values of $\lambda$,
terminating the evolution at a given value of $\Delta \phi$, which we take to correspond
to the end of the expansion phase.  We take
the time at which the expansion ceases to be $t_{cyc}$ in
Eq. (\ref{f}).  This allows us to plot $f$ as a function of
$\Delta \phi$ and $\lambda$, as shown in Fig. 1.
\begin{figure}[tbh]
\centerline{\epsfxsize=4truein\epsffile{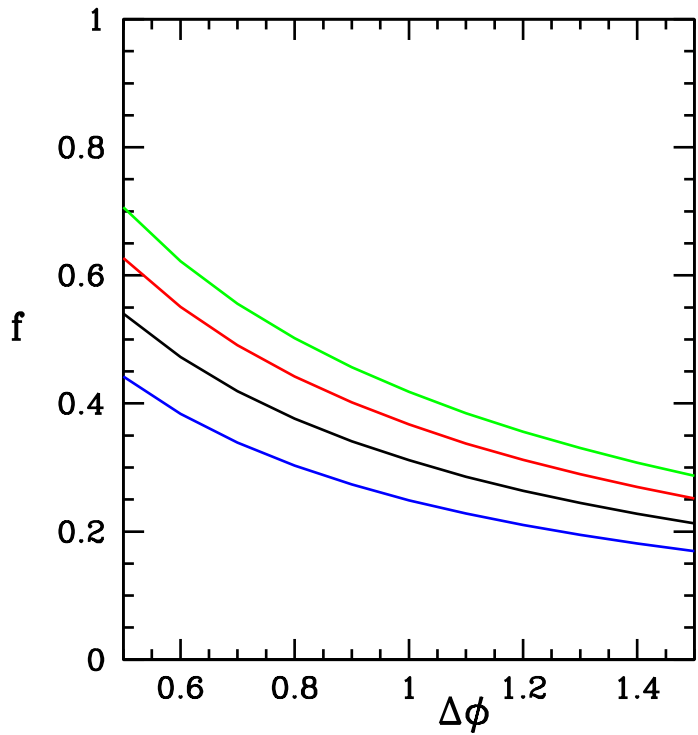}}
\caption{The coincidence fraction $f$, defined to be the fraction of the time that the universe spends
in a state in
which the matter and dark energy densities are within an order of magnitude of each other, for the IS
model driven by an exponential potential $V = V_0\exp(-\lambda \phi)$, as a function
of the total distance traversed by the scalar field, $\Delta \phi$, for (top to bottom),
$\lambda = 0.6$ (green), $\lambda = 0.5$ (red), $\lambda = 0.4$ (black), $\lambda = 0.3$ (blue).}
\end{figure}
As expected, smaller values
of $\Delta \phi$ and larger values of $\lambda$, both
favored by the Swampland conjectures, result in a shorter cycle time, increasing the likelihood
of finding ourselves in a ``coincidental" state today.  The precise value of $f$ depends on the (unknown)
Swampland limits (Eqs. \ref{swamp1} and \ref{swamp2}).  If, for example, we take $d \approx
1$ and $c \approx 0.6$, we find that $f > 0.42$, a reasonable solution
to the coincidence problem.

Of course, this is only one of many possible forms for $V(\phi)$ in the IS model, but it gives
a good qualitative estimate of the way that the coincidence fraction is likely to depend
on the Swampland parameters.  It might seem intellectually perverse to invoke the Swampland
conjectures in the context of the IS model, as one of the most interesting aspects of the IS model
is that it corresponds to a universe in which one never needs to invoke quantum effects to describe
the evolution, while the Swampland conjectures arise from attempts to model quantum gravity.
Nonetheless, there is no obvious reason to believe that the two ideas are incompatible.

Finally, note that this paper makes two separate claims, with two very different degrees of
disputability.  The argument that the IS model can help to resolve the coincidence problem is quite
strong. Given a value for $t_{cyc}$, one can calculate the exact fraction of time that we would
expect to find ourselves in a coincidental state. If this is a substantial fraction of unity,
then the IS model can be taken as a plausible solution to the coincidence problem.  Our
second argument, that the Swampland conjectures provide an upper bound on $t_{cyc}$,
is on shakier ground, if only because the Swampland conjectures are, in fact, conjectures,
and at this point lack specific values for the bounds specified in Eqs. (\ref{swamp1}) and
(\ref{swamp2}).  However, the first argument does not rely on the second.  If one
could find some other way to place an upper bound on $t_{cyc}$ in the IS model,
then the coincidence problem could be ameloriated without recourse to the Swampland conjectures.

\section*{Acknowledgments}

R.J.S. was supported in part by the Department of Energy (DE-SC0019207).

\end{document}